\documentclass[a4paper,fleqn,usenatbib]{mnras}

\usepackage{newtxtext,newtxmath}

\usepackage[T1]{fontenc}
\usepackage{ae,aecompl}

\usepackage{graphicx}	
\usepackage{amsmath}	
\usepackage{amssymb}	

\usepackage{lscape}
\usepackage{ulem}
\usepackage{url}
\usepackage{subfig}

\newcommand\dz{$\Delta z$}

\newcommand\mm{$\mu$m}
\newcommand\herschel{\textit{Herschel}}
\newcommand\xid{{\sc {xid+}}}

\newcommand\dmpc{$\Delta$Mpc}

\title[Blended \herschel~sources are LOS projections]{Cosmic happenstance: 24-\mm~selected, multi-component {\it Herschel}\thanks{{\it Herschel} is an ESA space observatory with science instruments provided by European-led Principal Investigator consortia and with important participation from NASA.} sources are line-of-sight projections.}

\author[Scudder et al.] {Jillian M. Scudder$^{1,2}$\thanks{Jillian.Scudder@oberlin.edu}, Seb Oliver$^{2}$, Peter D. Hurley$^{2}$, Julie L. Wardlow$^{3}$, \newauthor Lingyu Wang$^{4,5}$, Duncan Farrah$^{6, 7}$ \\
$^1$Department of Physics and Astronomy, Oberlin College, Oberlin, Ohio, 44074, USA\\
$^2$ Astronomy Centre, Department of Physics \& Astronomy, University of Sussex, Brighton, BN1 9QH, England\\
$^3 $Department of Physics, Centre for Extragalactic Astronomy, Durham University, South Road, Durham DH1 3LE, England\\
$^4$  SRON Netherlands Institute for Space Research, Landleven 12, 9747 AD, Groningen, The Netherlands\\
$^5$ Kapteyn Astronomical Institute, University of Groningen, Postbus 800, 9700 AV, Groningen, The Netherlands\\
$^6$ Department of Physics and Astronomy, University of Hawaii, 2505 Correa Road, Honolulu, HI 96822, USA\\
$^7$ Institute for Astronomy, 2680 Woodlawn Drive, University of Hawaii, Honolulu, HI 96822, USA
}

\date{Accepted XXX. Received YYY; in original form ZZZ}

\pubyear{2018}

\begin{document}
\label{firstpage}
\pagerange{\pageref{firstpage}--\pageref{lastpage}}
\maketitle

\begin{abstract}
In this paper, we investigate the physical associations between blended far-infrared (FIR)-emitting galaxies, in order to identify the level of line-of-sight projection contamination in the single-dish \herschel~data. 
Building on previous work, and as part of the Herschel Extragalactic Legacy Project (HELP), we identify a sample of galaxies in the COSMOS field which are found to be both FIR-bright (typically $\sim 15$ mJy) and blended within the \herschel~250 \mm~beam.
We identify a spectroscopic or photometric redshift for each FIR-bright source. We conduct a joint probability distribution analysis on the redshift probability density functions to determine the fraction of the FIR sources with multiple FIR-bright counterparts which are likely to be found at consistent (\dz $< 0.01$) redshifts. We find that only 3 (0.4 per cent) of the pair permutations between counterparts are $>50$ per cent likely to be at consistent  redshifts. A majority of counterparts (72 per cent) have no overlap in their redshift probability distributions whatsoever. This is in good agreement with the results of recent simulations, which indicate that single-dish observations of the FIR sky should be strongly contaminated by line of sight projection effects.
  We conclude that for our sample of 3.6- and 24-\mm~selected, FIR-bright objects in the COSMOS field, the overwhelming majority of multi-component FIR systems are line of sight projections within the 18.1 arcsec \herschel~beam, rather than physical associations.

\end{abstract}

\begin{keywords}
galaxies: high redshift -- galaxies: statistics -- galaxies: star formation -- galaxies: starburst
\end{keywords}

\section{Introduction}
\label{sec:intro}
As surveys of the galaxy population have evolved, we have become sensitive to an increasingly diverse population of galaxies, and continue to push to higher redshifts. Of particular interest are those galaxy sub-populations that challenge the predictions of theoretical models.
The population of luminous galaxies in the far infra-red (FIR) and sub-mm \citep[e.g.,][]{Smail1997, Hughes1998, Barger1998} has posed a particular challenge for our current understanding of galaxy evolution.

These galaxies, originally discovered in blind surveys in the sub-mm, have been subjected to a number of follow-up programs, which have determined that these galaxies are typically found at high redshift \citep[e.g., ][]{Smail2000, Smail2002, Chapman2005}, and that their luminosity at these wavelengths is due to the presence of large amounts of heated dust. This dust is presumably heated by the presence of significant star formation within the galaxy; the dust is absorbing the UV radiation from massive young stars, and reradiating it at longer wavelengths. For reviews, see \citet{Blain2002} and \citet{Casey2014}. 

These galaxies have posed a significant challenge to theoretical models of galaxy formation, as their luminosity implies astoundingly high star formation rates at very early times in the Universe. Models have invoked a number of potential solutions in order to drive these star formation rates, including merger-induced star formation \citep[e.g.,][]{Narayanan2010, Dave2010}, a variable (top-heavy) IMF \cite[e.g.,][]{Baugh2005}, large-scale disk fragmentation \citep[e.g.,][]{Immeli2004}, pristine gas infall \citep[e.g.,][]{Dekel2009, Narayanan2015}, or some combination thereof.

Testing the theoretical explanations for such elevated SFRs has been challenging, as observations of the sub-mm/FIR luminous galaxy population were historically limited to single dish facilities with very limited resolution. The typical single-dish facility in the sub-mm has a full width half maximum of about 20 arcsec.
 This low resolution presents several problems, the most severe of which is that it renders accurate counterpart identification at shorter wavelengths (where the resolution is improved) difficult. Within the beam, it is not uncommon to find a number of potential optical counterparts \citep[e.g.,][]{Hughes1998, Downes1999, Dunlop2004, Ivison2007, Clements2008},
 and so the identification of the most appropriate counterpart is not straightforward.

This counterpart identification is further complicated by the anticipated difficulty of detecting the FIR flux-emitting galaxy in the optical at all, considering their high redshifts. Because the source of the FIR emission is the joint presence of star formation and large quantities of dust, the counterparts which are most likely to be strongly contributing to the observed sub-mm flux are also likely to be heavily dust-obscured. Much of the optical light will have been absorbed by the very dust that renders them so luminous in the FIR, making the optical colours of these objects extremely red. Furthermore, at the highest redshifts, optical counterparts become increasingly faint. FIR sources, on the other hand, benefit from a negative k-correction, which keeps them visible as bright sources over a very wide redshift range \citep{Blain2002}. For example, a $5\times10^{12}$L$_{\odot}$ galaxy, observed at $\sim$250 \mm, is visible as a >10 mJy source out to $z\approx2.5$ \citep{Blain2002}.

In spite of these challenges, for some of the earliest-identified sub-mm sources found in regions of the sky with very deep optical data, redshifts were obtained \citep{Ivison1998, Hughes1998, Downes1999, Ivison2000, Frayer2000, Chapman2005}. As methods advanced, a number of sub-mm galaxies (SMGs) were successfully targeted with optical telescopes, based on prior radio counterpart identifications \citep[e.g.,][]{Ivison2002}.
Further observations could map the gas content (typically CO) of a handful of these systems \citep{Frayer1998, Frayer1999, Ledlow2002, Neri2003, Greve2005, Hainline2006, Tacconi2006, Tacconi2008}. Much of the work on the earliest known sources concluded that due to the clumpy/irregular morphologies or high gas densities, these systems were likely to be late-stage mergers \citep{Tacconi2006, Tacconi2008, Ivison2002, Smail2003, Engel2010, Younger2010,Zamojski2011, Menendez2013, Wiklind2014, Chen2015}, though cf \citet{Swinbank2011}, which found evidence for a rotating disk.

This proposed merger-induced origin of sub-mm galaxies was also used to explain the higher than expected levels of clustering found in the SMG population \citep[e.g.,][]{Blain2004, Farrah2006, Amblard2011, Cowley2016}. Studies also found that there were a higher than expected number of radio counterparts in close proximity to FIR sources; a statistical argument was made arguing for the unlikelihood of this arrangement at random \citep{Ivison2007}.
Merger-induced star formation, which had been predicted by some theoretical models \citep[e.g.,][]{Baugh2005, Swinbank2008, Dekel2009, Narayanan2010, Dave2010}, could be invoked to explain the unusually high levels of star formation \citep[e.g.,][]{RowanRobinson2017} within these systems. A detailed discussion of the origins of this merger-induced model of sub-mm galaxies is undertaken in the Appendix.

As sub-mm interferometric facilities, such as ALMA, have come online, we have begun to observe these sources at much higher resolution without needing to change wavelengths. High-resolution studies have typically found that some sub-mm sources tend to divide into a number of FIR-bright components \citep{Karim2013, Hodge2013, Simpson2016, Trakhtenbrot2016}. In these cases, the star formation rates previously attributed to a single galaxy should be divided amongst multiple components, reducing the extremity of star formation in each individual object, though if these objects are interacting, the SFR of the system remains elevated. \citet{Scudder2016} similarly found that single-dish FIR sources are also best reproduced by multiple FIR-bright components, using a statistical method \citep{Hurley2017}. 

The interpretation of the single-dish flux therefore remains unclear. If these counterparts are all at the same redshift, the unresolved FIR flux traces the SFR of a physical system instead of a single galaxy. However, the theoretical challenge remains if the components are physically associated. 
If these multiple components are all part of interactions, theoretical models must still produce starbursts at early times in order to replicate the observations, which continues to be a significant challenge. Alternately, if these multiple components are physically unrelated, appearing close on the sky by virtue of line of sight projection effects and a broad range of redshift visibilities, then the theoretical models do not need to strain as hard to produce massive bursts of star formation in the very early Universe. Indeed, \citet{Efstathiou2003} predicted that many of these observed sub-mm galaxies may be closer to normal star forming galaxies than extreme starbursts. Secular disk fragmentation, also suggested as a pathway to generate sub-mm galaxies \citep[e.g.,][]{Immeli2004, Bournaud2009, Dekel2009} may be a viable explanation for the irregular morphologies observed.

As theoretical models become more complex, simulations have begun to assess the frequency of single-dish observations blending sources together. A number of simulations now suggest that source blending should be significant, both for physically associated galaxies \citep{Hayward2013a} and for galaxies which are entirely unrelated \citep{Hayward2011, Hayward2013, Cowley2015, Munoz2015, Cowley2016, Bethermin2017}. The fraction of totally unrelated sources varies between simulations, but has been estimated to be in the range of 50 per cent to as high as 70 per cent \citep{Hayward2013, Cowley2015, Bethermin2017}.

In this work, we build upon previous work which identified a sample of FIR-bright counterparts to a sample of \herschel~sources in the COSMOS field \citep{Scudder2016}. We identify the spectroscopic or photometric redshifts associated with these FIR-luminous counterparts and test directly, on a statistical sample, whether these multiple-component systems are likely to be physically associated or are simply line-of-sight projections. The results of this work will give a framework for understanding the meaning of a FIR-blended source more broadly.

In Section \ref{sec:sample} we describe the selection of our FIR bright counterpart sample. In Section \ref{sec:analysis} we describe our data analysis. In Section \ref{sec:discussion} we discuss the implications of our work in the context of the literature. In Section \ref{sec:conclusions} we present our conclusions. The full set of figures for all FIR sources within our sample is available online\footnote{Figures are available at this http url: \href{https://github.com/jmscudder/Redshift-figs}{https://github.com/jmscudder/Redshift-figs}}.
Throughout this work, we assume WMAP 9 cosmology \citep{WMAP9}\footnote{This cosmology is $\Omega_M = 0.282, \Omega_{\Lambda}=0.718, H_0=69.7$ km s$^{-1} $ Mpc$^{-1}$}.

\section{Sample Selection}
\label{sec:sample}
In this work we build upon the work of \citet{Scudder2016}, and use the sample defined therein. For a full description of the sample selection, we refer the reader to that work. For clarity, we also provide a brief description of our selection criteria here.

\subsection{The \citet{Scudder2016} sample}
\citet{Scudder2016} defines its sample within the COSMOS field region \citep{Scoville2007} because of the strong multi-wavelength coverage in that field. In particular, \citet{Scudder2016} makes use of the 250-\mm~coverage of the COSMOS region by the Spectral and Photometric Imaging Receiver \citep[SPIRE,][]{Griffin2010} instrument onboard the \textit{Herschel Space Observatory} \citep{Pilbratt2010}. These observations were part of the \herschel~Multi-tiered Extragalactic Survey \citep[HerMES;][]{Oliver2012}.  \citet{Scudder2016} also made use of pre-existing 3.6 and 24-\mm~catalogues in this field. 
We summarise the data used in Table \ref{tab:datasrc}.

\begin{table}

\caption{A summary of the data used in this work.}
\centering
\resizebox{\columnwidth}{!}{%
\begin{tabular}{|l|c|c|r|}
Wavelength & Flux density limit & Facility & Data Source\\
\hline
250 \mm & $\geq$ 30 mJy & \herschel & \citet{Levenson2010}\\
&&& \citet{Wang2014a}\\
24 \mm & $\geq$ 150 $\mu$Jy & \textit{Spitzer}& \citet{LeFloch2009}\\
3.6 \mm & $\geq$ 0.9 $\mu$Jy & \textit{Spitzer}& \citet{Sanders2007}\\
\hline
\end{tabular}
}
\label{tab:datasrc}
\end{table}%

Each  250-\mm~source above a flux density threshold of 30 mJy was crossmatched with the 3.6- and 24-\mm~catalogues. FIR sources were preserved in the sample if they had $\geq1$ detection at 3.6 \mm~and 24 \mm~within 18.1 arcsec (the FWHM of \herschel~at this wavelength) of the 250-\mm~catalogue position. The  3.6-\mm~and 24-\mm~wavelength detections were permitted to spatially overlap, and there was no requirement for $>1$ detection in each band. 
\citet{Scudder2016} identified 360 such FIR sources, with a median number of 14 possible multi-wavelength (3.6- and/or 24-\mm) counterparts\footnote{A point of linguistic clarity: in this work we refer to the FIR detection as the `object' or `source', with the positions of the higher-resolution 3.6- and 24-\mm~ detections as `counterparts' to the FIR detection.} per 250 \mm~detection.

The \xid~software \citep{Hurley2017} was used to identify the most probable distribution of FIR flux amongst these potential counterparts. The methodology and tests of this tool on simulated data are fully explored in \citet{Hurley2017}. Briefly, \xid~is a Bayesian inference tool that uses the positions\footnote{In this iteration of \xid, the only prior information used is that of the positions. The software is under active development (available at \href{https://github.com/H-E-L-P/XID_plus}{https://github.com/H-E-L-P/XID\_plus}) and the ability to use flux priors will soon be available.}  of known objects to determine the most likely distribution of flux between those known objects, so that the input map (in this case, the 250-\mm~map) is best reproduced. 
\citet{Hurley2017} shows that \xid~both accurately recovers fluxes in a synthetic map of the COSMOS region, and produces accurate estimations of its errors. As the output of \xid~is a full posterior distribution function, any non-gaussianities in the flux solutions, or strong degeneracies between solutions (e.g., in the case that two sources are too close together for the software to provide a preferred solution) are preserved. The full set of flux solutions and correlations between sources for the \citet{Scudder2016} sample are available online\footnote{Flux solutions and intercorrelation figures are available for the 360 FIR sources at the following URL: \href{http://jmscudder.github.io/XID-figures}{http://jmscudder.github.io/XID-figures}.}.
In \citet{Scudder2016}, all known 3.6-\mm~sources within a 180 arcsec by 180 arcsec region surrounding each FIR source of interest are used in the fitting procedure, in order to avoid poor fits due to other bright sources near the FIR source of interest.

\subsection{The FIR-bright subsample}

In this work, we are interested in investigating the redshifts of the FIR-bright population. As the \xid~analysis provides a full probability density function (PDF) of possible flux solutions, defining a level of FIR-brightness is somewhat arbitrary. An absolute flux threshold preferentially selects counterparts from brighter FIR sources, so to avoid this bias, we use the fraction of the total FIR flux assigned to a given component as a more scalable method of selecting sources that significantly contribute to the FIR flux observed in the map. For each \xid~flux solution, we calculate the flux ratio of a given counterpart relative to the sum of all contributing sources. This builds up a PDF for the flux ratios. Any counterparts that have a median flux ratio greater than ten per cent are flagged as significantly contributing (henceforth FIR-bright). 

We note that for very low flux ($\sim$few mJy) solutions, the final PDF of flux solutions is strongly non-gaussian, and the median is not a good estimator of the typical flux solution.  However, once the flux solutions rise above a few mJy, the median is typically a good estimator of the flux solution distribution \citep{Hurley2017}. In \citet{Scudder2016} we chose a threshold of ten per cent of the total flux as the limit for significantly contributing. For consistency, we retain that definition in this work. As our lowest FIR flux source is 30 mJy, the smallest possible flux which could pass the 10 per cent threshold is 3 mJy. The median flux in our FIR-bright counterpart sample is $\sim$15 mJy, with very few (2.6 per cent) components with fluxes < 5 mJy. Our median values should therefore be reliable estimators of the posterior distribution. As above, the complete set of figures which show median estimates of all flux solutions from \citet{Scudder2016} are available online.

\begin{figure}
\begin{center}
\includegraphics[width=250px]{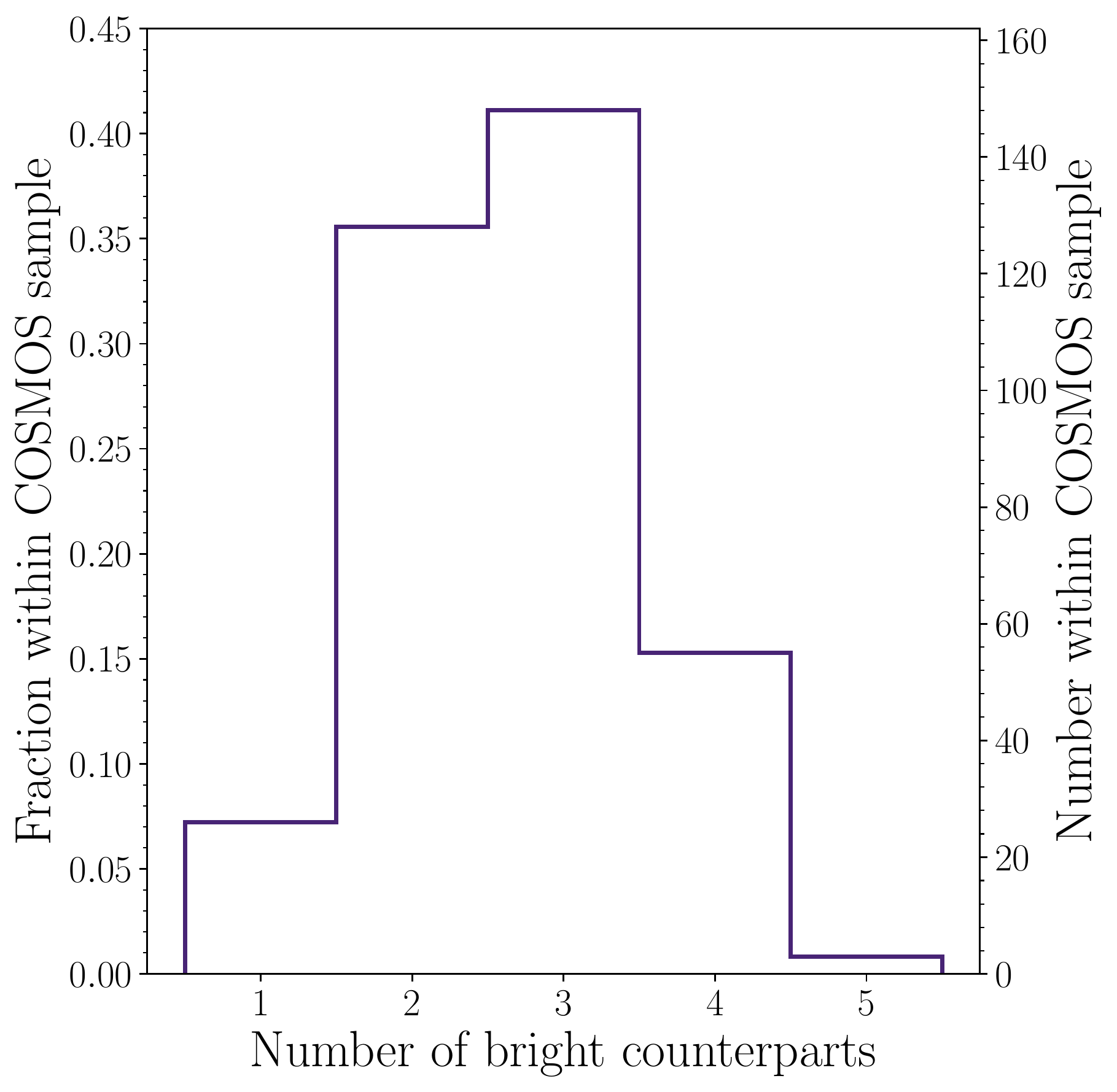}
\caption{A histogram showing the number of FIR-bright counterparts per FIR source, per the {\sc{xid+}} analysis. FIR-bright is defined as a counterpart which contributes $>10$ per cent of the total FIR flux. The left hand vertical axis shows the fraction of the total sample in each bin, and the right hand vertical axis gives this value as a number of FIR objects. The majority of FIR sources have 2-3 FIR-bright counterparts. }
\label{fig:counterpartfrac}
\end{center}
\end{figure}

In Figure \ref{fig:counterpartfrac}, we show a histogram of the number of FIR-bright components per FIR object for the full sample of 360 FIR objects from \citet{Scudder2016}. Along the left vertical axis, we show the fraction of the 360 FIR objects which have the given number of bright counterparts; on the right, we translate this into a raw number of FIR objects.  As reported in \citet{Scudder2016}, the vast majority of FIR objects have more than one bright component. However, approximately 7 per cent (26 individual objects) of the 360 FIR objects are comprised of a single component only. These single-component objects are those FIR detections which are best explained via a single luminous counterpart.

We wish to examine the redshifts of the objects which are blends of multiple FIR-bright components, and so we exclude these single-counterpart objects for the remainder of this work. However, we note that since these objects do not make up a significant fraction of the overall sample, this is not a major reduction in sample size\footnote{The exclusion of these 26 objects may influence mean statistics, though we emphasize the small number of sources.}. 
Once this criterion has been put in place, our sample contains 334 FIR objects with more than one bright counterpart, with a total of 935 FIR-bright counterparts within these fields.

\subsection{Redshift identification}
We now identify the redshift of each FIR-bright counterpart. The COSMOS field benefits from extensive spectroscopic and photometric redshift availability. We use the \citet{Davies2015} catalogue of spectroscopic redshifts, which collects and reprocesses the zCOSMOS spectra \citep{Lilly2009}, along with spectra from VVDS \citep{LeFevre2013}, PRIMUS \citep{Cool2013}, and SDSS \citep{Ahn2014}. 

For photometric redshifts, we use the \citet{Laigle2016} catalogue of photometric redshifts. The \citet{Laigle2016} photometric catalogue is extremely well calibrated to spectroscopic redshifts, and has a precision relative to the COSMOS spectroscopic catalogue \citep{Lilly2009} of $\frac{|z_p - z_s|}{1\textrm{+}z_s}$=0.007, with a catastrophic failure rate of 0.5 per cent \citep{Laigle2016}. Above a redshift of 3, the \citet{Laigle2016} catalogue maintains a redshift precision of $\frac{|z_p - z_s|}{1\textrm{+}z_s}$=0.021, with a failure rate of 13.2 per cent. 

We implement basic quality filters on the \citet{Laigle2016} catalogue and on our spectroscopic catalogue from \citet{Davies2015}, which are summarized in Table \ref{tab:zqual}. We require that any redshift solution is above zero ($z$>0). 
The photometric catalogue is additionally required to be outside of a star mask region, {\sc{star\_flag}}=$<1$. This combination of flags is similar to the recommended quality flag {\sc{flag\_peter}}=0, but not identical, as a number of our sources lie outside the main COSMOS region, which is masked by both {\sc{flag\_cosmos}}=1 and {\sc{flag\_peter}} =0.
We also require {\sc{z\_use}}$<3$, which limits the spectroscopic redshifts to either high resolution spectra (e.g., from zCOSMOS) or where reliable redshifts were estimated from PRIMUS \citep{Davies2015}.

\begin{table}
\caption{A summary of the quality control flags imposed on the spectroscopic and photometric redshift catalogues.}
\begin{center}
\begin{tabular}{|l|r|}
Photometric Catalogue & Spectroscopic Catalogue\\
\hline
$z > 0$ & $z>0$\\
{\sc{star\_flag}}=$<1$ &  {\sc{z\_use}}$<3$ \\
\hline
\end{tabular}
\end{center}
\label{tab:zqual}
\end{table}%

In order to identify the redshift associated with each of our counterparts, we search a 2 arcsec radius surrounding the shorter-wavelength (3.6- or 24-\mm) counterpart location, and identify the nearest spectroscopic redshift and the nearest photometric redshift to that.
 In the vast majority of cases, only one redshift type exists, and that redshift, be it spectroscopic or photometric, is accepted as the redshift of the shorter-wavelength counterpart. In the case where both spectroscopic and photometric redshifts exist, the closer spatial identification is accepted.  

If there are no spectroscopic or photometric redshift matches within 2 arcseconds, the FIR-bright counterpart is flagged as having no identifiable redshift, and excluded from the analysis that follows. Some FIR objects have no redshifts for any of their 3.6 or 24-\mm~counterparts. We determined that the vast majority of the redshift identification failures are due to contamination of the optical imaging by the presence of a bright star. The bright star flux renders the optical imaging sufficiently unreliable that photometric redshifts were not calculated. 

If all counterparts for a given FIR object fail the redshift identification process (true for 16/334 of our FIR objects, or 4.8 per cent), the FIR object is excluded from the analysis that follows. 
In some cases, only one counterpart associated with a FIR object has a redshift. In this case, no pairwise redshift comparison can be undertaken and the FIR object must be discarded from our analysis. As long as there remains $>1$ counterpart with a redshift identified, the FIR object is preserved in the analysis. We discuss the possible impact of this loss of individual counterparts to the results of our analysis in Section \ref{sec:discussion}, but find that this should not significantly affect the results presented here.

\begin{table}
\caption{Description of the distribution of redshift identifications. The top rows indicates the number of total FIR-bright components, and the number of those components which had identifiable redshifts. The bottom rows subdivide the counterparts which were missing redshifts.}
\begin{center}
\begin{tabular}{|l|r|}
Total components: & 935 \\
 Counterparts with identified redshifts: & 806\\
\hline
Counterparts without redshifts: & 129\\
\hline
0 counterpart redshifts remaining per FIR object: & 37/129\\
$ 1$ counterpart redshift remaining per FIR object: & 49/129\\
$\geq 2$ counterpart redshifts remaining per FIR object: & 43/129\\
\hline
\end{tabular}
\end{center}
\label{tab:missingz}
\end{table}%

806 of 935 counterparts have redshift identifications (86 per cent). We break down the redshift identification results in Table \ref{tab:missingz}. Of those that are missing, 37 counterparts are associated with one of the 16 FIR objects that have no redshift identifications at all. These blank regions reduce our FIR object count from 334 to 318. The remaining 92 redshift-unidentified counterparts have at least one other redshift within the field.  49 of the 92 unmatched counterparts have only one counterpart with a redshift identifications associated with a specific FIR source, which results in the loss of 38 FIR fields (along with 38 redshift identifications) from our analysis. Our final sample contains 280 FIR objects, with 768 associated counterparts, which have had $\geq 2$ redshifts successfully identified. 95 (12.4 per cent) of the final sample has spectroscopic redshifts; 673 (87.6 per cent) counterparts have photometric redshifts.

\section{Analysis}
\label{sec:analysis}
\begin{figure}
\subfloat{\includegraphics[width=250px]{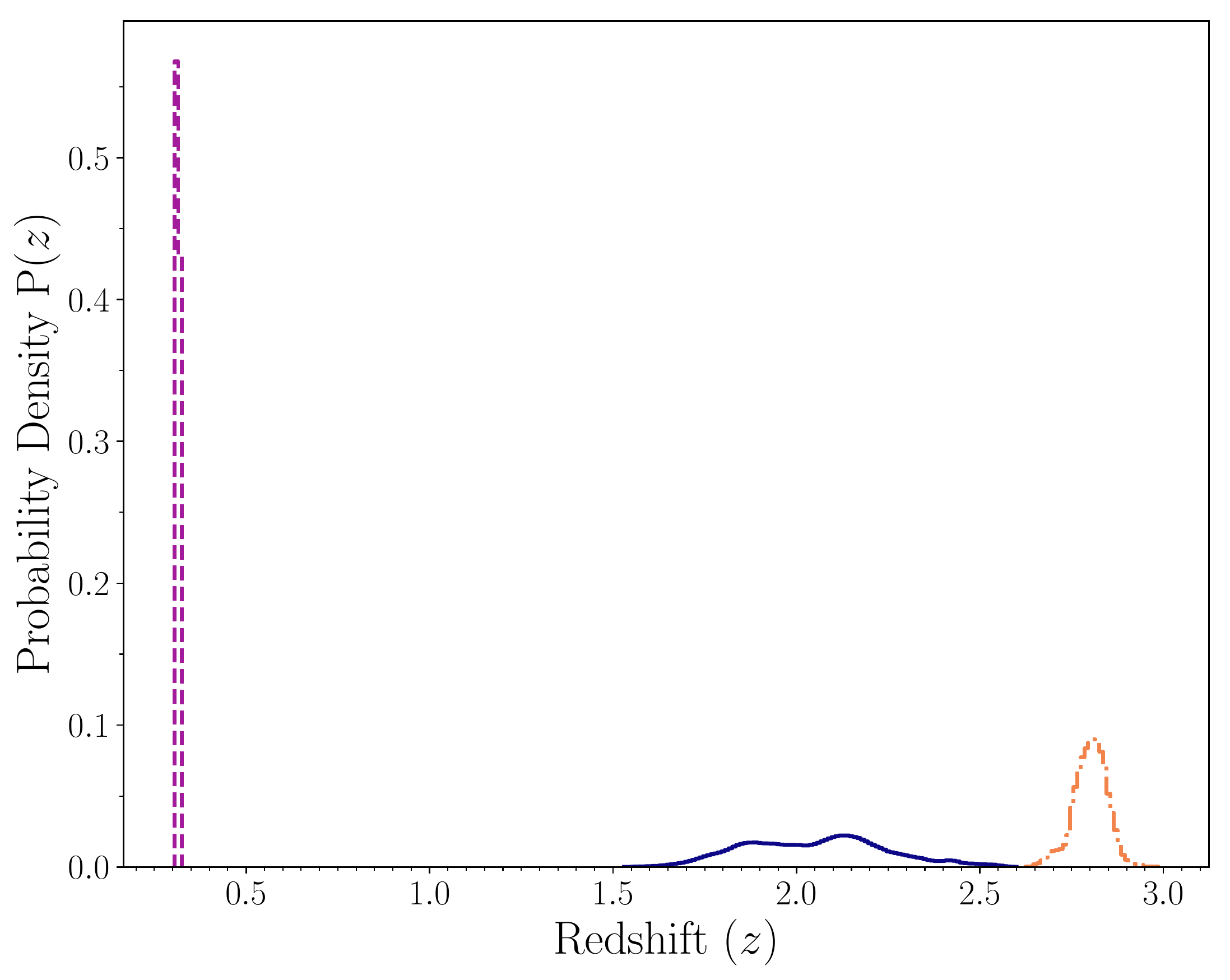}}\\[-3ex]
\subfloat{\includegraphics[width=250px]{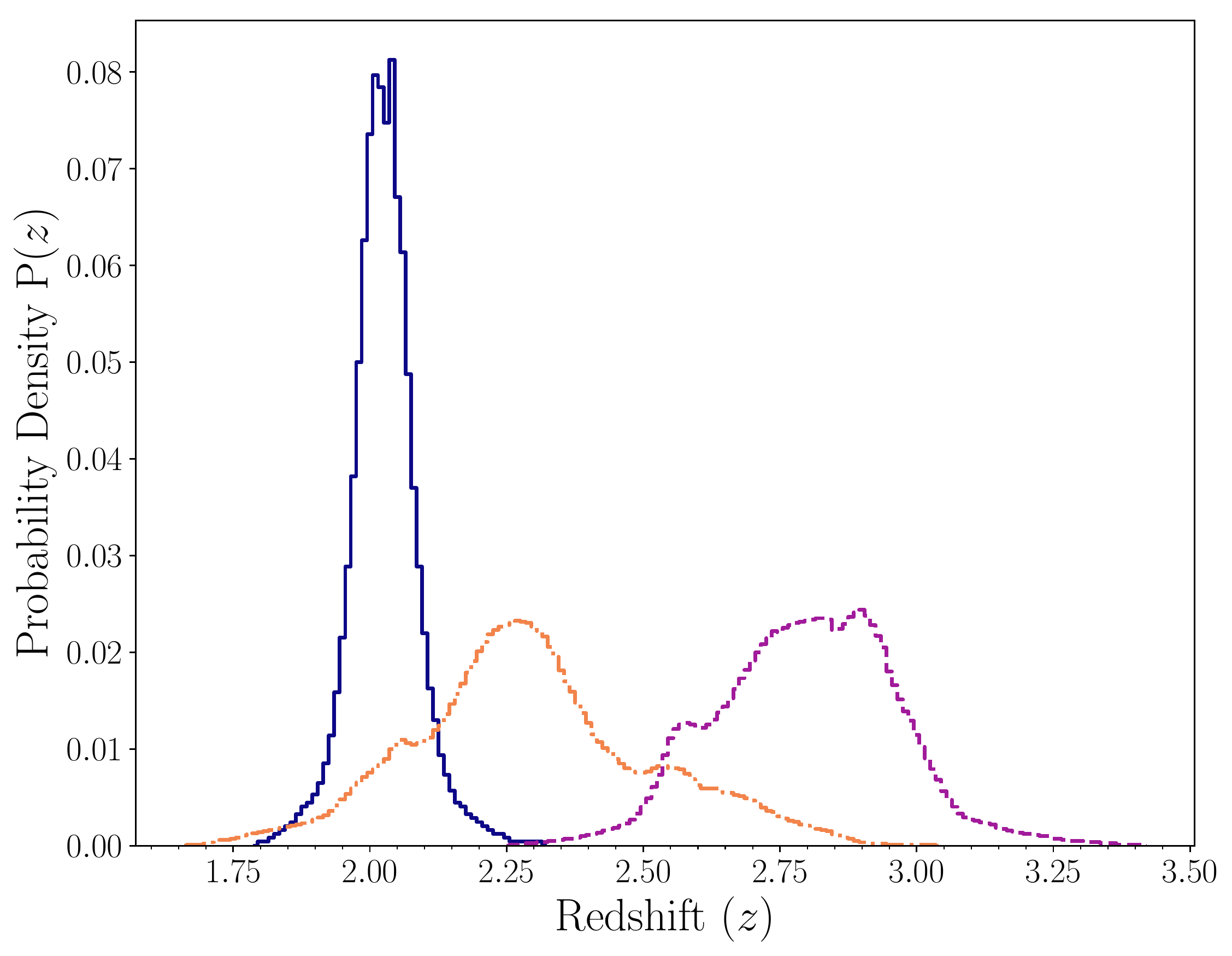}}
\caption{The redshift probability distribution functions of two sample FIR sources, with 3 FIR-bright counterparts each. Very broad distributions indicate less well-confined photometric redshift distributions, whereas narrow peaks indicate spectroscopic observations or (as here) very good photometric redshift solutions. The cumulative probability has been normalized to one. We calculate the joint probability distribution for all combinations of pairs in these systems. In these cases, there are 3 unique permutations: [purple dashed]--[blue solid], [purple dashed]--[orange dot-dashed], and [orange dot-dashed]--[blue solid]. The top panel shows ID 3422. The redshift PDFs do not overlap within a tolerance of $z$=0.01. In the bottom panel, we show ID 9007, where there is significant overlap between the PDFs.\label{fig:zpdf}}
\end{figure}

We make use of the full photometric PDFs, which are discretized in redshift bins\footnote{At a redshift of 1.0, the median of our sample, \dz=0.01 corresponds to a change in luminosity distance of $\sim$83 Mpc. \dz=0.05 corresponds to $\sim$415 Mpc.} of \dz=0.01, from the \citet{Laigle2016} catalogue for all counterparts with photometric redshifts. 
We renormalize the redshift PDFs such that the cumulative probability is equal to 1. If a source has been matched to a spectroscopic redshift, we assume that its probability density function has 100 per cent probability of being found within a single \dz=0.01 redshift bin. In Figure \ref{fig:zpdf}, we show the PDFs of the redshifts for two randomly selected FIR sources (ID 3422, top panel; ID 9007, bottom panel). 3422 has one counterpart with a very well constrained photometric PDF at $z$ = $0.3$, plotted in a dashed purple line, which has non-zero probability contained within two \dz=0.01 bins. The other two counterparts' redshifts have much broader PDFs. 
In the bottom panel, we show an example where the counterparts' redshift PDFs are found to be in overlap. Here we have three moderately well constrained PDFs, with the counterpart plotted in orange dot-dashed in overlap with both other counterpart PDFs.
All other fields with redshift identifications for more than one component and are available online\footnote{These figures are available online at \href{https://github.com/jmscudder/Redshift-figs}{https://github.com/jmscudder/Redshift-figs}}.  Unless otherwise specified, we retain the discretization in redshift space at \dz=0.01. We note that while this discretization is smaller than the typical error at $z>$3.0, the fraction of our sources which are found above a redshift of 3.0 is small, and we have repeated our analysis with larger \dz~bin sizes and find our results are broadly unchanged.

\subsection{The redshift distribution of the FIR-bright sample}
\label{sec:zdist}
\begin{figure}
\begin{center}
\includegraphics[width=250px]{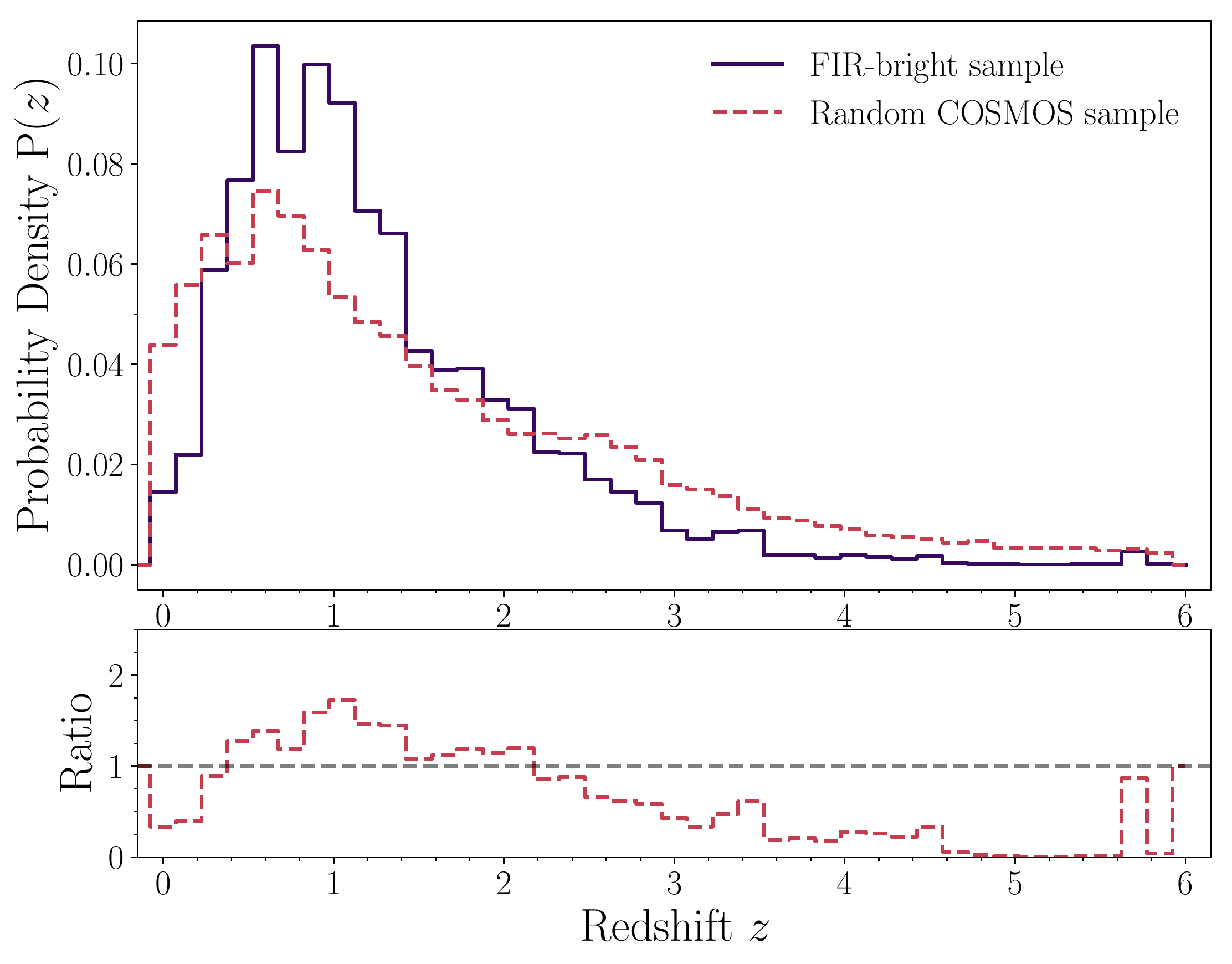}
\caption{The summed redshift probability for \textit{all} FIR-bright sources across all FIR objects, plotted as the purple solid line. We have expanded the bins to \dz=0.15. In a red dashed line, we show a random sampling of 5,000 redshifts from the photometric redshift catalogue. We see that both redshift distributions largely span the same range, but the COSMOS catalogue is less peaked at redshifts of $0.4<z<1.5$, and has a stronger tail to high redshifts. In the lower panel, we plot the ratio between the two curves: ${\left( \frac{FIR~bright}{random}\right) }$ at each $z$. This ratio shows explicitly the excess of probability in the FIR sample in the redshift range $0.4<z<1.5$.}
\label{fig:zpdfall}
\end{center}
\end{figure}

We first wish to determine whether there is any systematic difference in the redshift distribution of the FIR-bright counterparts and the COSMOS photometric catalogue from which it was drawn. We therefore build up the cumulative probability distribution across all FIR bright counterparts, effectively summing the counterpart PDFs shown in Figure \ref{fig:zpdf} across all FIR-bright counterparts, described by: 

\begin{equation}
\text{P(\it{z}) =} \sum_{c=0}^{c}{(P_{c}\text{(\it{z})})},
\label{eq:cumul}
\end{equation}

\noindent where $z$ is a given redshift solution, P($z$) is the cumulative probability density at a given redshift, $c$ is an index ranging from 0 to 752 (the number of counterparts), and P$_{c}\left(z\right)$ is the probability density for a given counterpart at the given redshift. The result of this operation\footnote{This analysis assumes that all PDFs within the \citet{Laigle2016} catalogue are independent of each other.} is shown as the purple solid line in Figure \ref{fig:zpdfall}, normalized so that the sum of the probability density is equal to 1. For clarity, in this figure we have expanded the bins to \dz=0.15.

To compare to the COSMOS sample, we randomly draw 5,000 galaxies from the quality-controlled COSMOS photometric catalogue, and compute their cumulative probability distributions in the same way. The only change to Equation \ref{eq:cumul} is that $c$ now ranges from 0 to 5,000. The cumulative probability density function for the random sampling is shown as the red dashed line in Figure \ref{fig:zpdfall}. It is immediately clear that while both samples cover a broad range of redshift space, the FIR-bright sample is more tightly clustered around a redshift of 1.0 than the randomly selected sample. The randomly selected photometric sources, by contrast, has a stronger tail out to both lower and higher redshifts. This peak at $z$=1.0 in the FIR-bright sample is not surprising, as it has been found that the 24 \mm~selected, 250-\mm~\herschel~sources (similar to the selection in this work) have a peak in their redshift distribution around this value, with previous work reporting the peak between $z$=0.85 \citep{Casey2012} and $z\sim1.0$  \citep{Bethermin2012}. Surveys selected at longer wavelengths, for instance an 850 \mm~selection, typically peak at a redshift of $z\approx2.5$ \citep[e.g.,][]{Chapman2005, Casey2012}

We plot the ratio between the two samples ${\left( \frac{FIR~bright}{random}\right) }$ in the bottom panel of Figure \ref{fig:zpdfall} in a dashed red line. This ratio shows the relative excess or deficiency in probability that a FIR-bright counterpart is found at a specific redshift, compared to the randomly selected sample, and more clearly demonstrates the differences between the two histograms. The FIR-bright counterpart sample is underrepresented at all redshifts relative to the COSMOS photometric catalogue except between redshifts of $0.4 \lesssim z \lesssim 2.0$. The peak of the FIR-bright sample at $z \approx 1.0$ can be seen as an excess of probability density of 1.5 times the random COSMOS catalogue.

\subsection{\dmpc~probability function}
We wish to estimate the redshift difference between FIR-bright components contributing to a given FIR object without losing valuable information from the PDFs. As Figure \ref{fig:zpdf} illustrates, the width of the redshift PDFs can vary significantly, and the broadest redshift solutions are not always well described by their medians. We must therefore estimate both the range in possible redshift differences (\dz), and the likelihood of each possible value of \dz. In order to address the large redshift ranges probed by this calculation, we convert redshift to a comoving distance.

For each FIR source, we identify all pair permutations between counterparts. In Figure \ref{fig:zpdf}, we have three unique pair permutations:  [blue dashed]--[purple solid], [blue dashed]--[green dot-dashed], and [green dot-dashed]--[purple solid]. For a given pairing, we denote the PDFs of the two counterparts as $P_a$ and $P_b$. We select the regions of redshift space over which P($z$)>0, and convert the remaining redshifts into comoving distances.
For each permitted comoving distance solution, $D_a$, in $P_a$, we subtract all possible distance solutions, $D_b$, in $P_b$. We store the absolute value of the difference:  $\Delta D_{ab} $~=~$ \left|D_a - D_b\right|$.

In order to retain the probability information, for each \dmpc~solution we weight by the probability of the two distance solutions $D_a$ \& $D_b$, described by:

\begin{equation}
P\left(\Delta \text{Mpc} \text{=}\Delta D_{ab}\right) $~=~$ P_a\left({D_{a}}\right) \times P_b \left(D_{b}\right)
\label{eq:dzjoint}
\end{equation}

\begin{figure}
\subfloat{\includegraphics[width=250px]{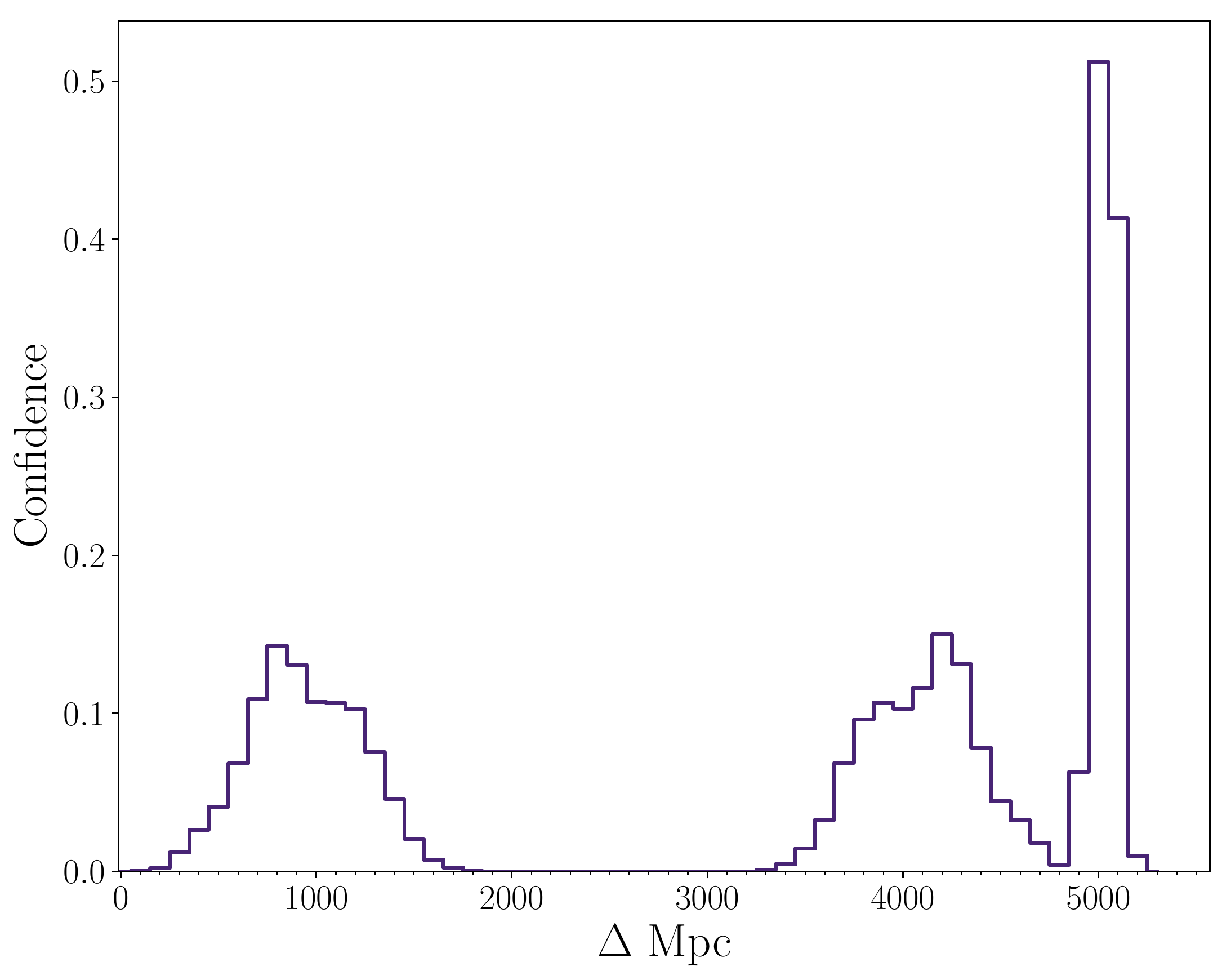}}\\[-3ex]
\subfloat{\includegraphics[width=250px]{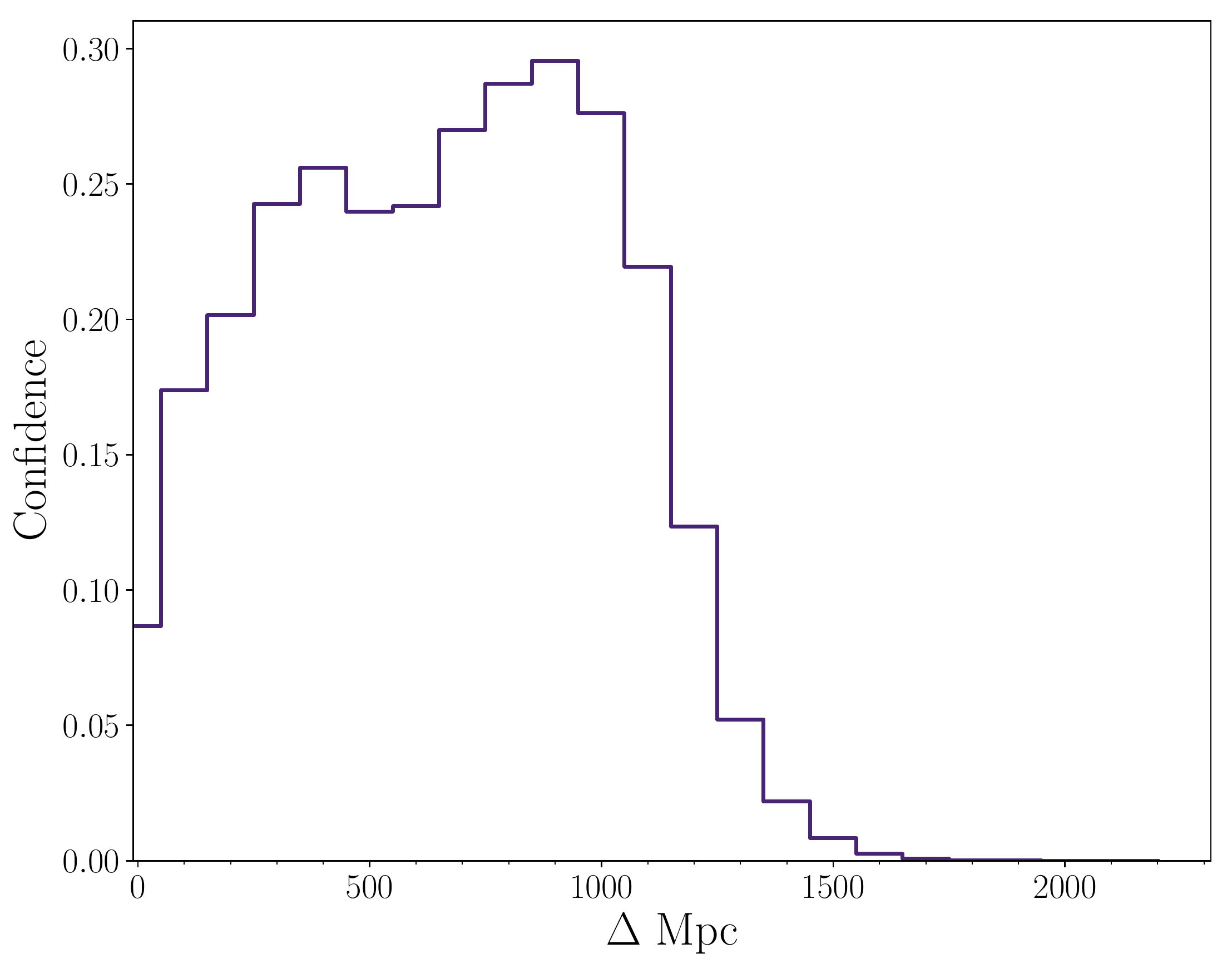}}
\caption{All possible solutions for $\Delta$Mpc, weighted by their probability, for the same set of counterparts as Figure \ref{fig:zpdf}. In the top panel we again show ID 3422, and in the bottom panel, ID 9007. For each unique pair of counterparts, we calculate all possible differences in comoving distance between it and the other bright components, given the range of solutions in the PDF. We then weight the differences by the likelihood of the redshift solutions $P\left(z\right)$. The shape of the original PDFs are reflected in the top panel, though somewhat horizontally compressed due to the conversion into Mpc that has been applied. In the bottom panel, we do not see the same distinct peaks, as the original PDFs are significantly in overlap. The cumulative probability in these figures sums to the number of counterparts (i.e., 3.0 for both panels).}
\label{fig:dzpdf}
\end{figure}

Duplicate \dmpc~values (of which there should be many, as adjacent bins provide similar \dmpc) have their probabilities summed together for the final analysis.  Once all \dmpc~and $P\left(\Delta \text{Mpc}\text{=}\Delta D_{ab}\right)$ values have been determined for a given pair, this process is repeated for all pair permutations that exist for that FIR object. We demonstrate the output of this method by showing the results for ID 3422 and ID 9007 in the top and bottom panels of Figure \ref{fig:dzpdf}, respectively. The full set of these figures for all FIR objects is also available online\footnote{\href{https://github.com/jmscudder/Redshift-figs}{https://github.com/jmscudder/Redshift-figs}}.

In the top panel, the shapes of the original PDFs are reflected in the shape of the histogram. The stronger peak at $\Delta$Mpc=5000 reflects the difference between the purple dashed and orange dot-dashed PDFs in the top panel of Figure \ref{fig:zpdf}. The middle peak reflects the differences between the purple dashed and blue solid peaks, and the leftmost peak reflects the difference between the blue solid and orange dot-dashed peaks in Figure \ref{fig:zpdf}. In the bottom panel, we see a much broader distribution of \dmpc~values for ID 9007. As the original PDFs were much more strongly in overlap, a continuous distribution of \dmpc~values is to be expected.

\begin{figure}
\begin{center}
\includegraphics[width=240px]{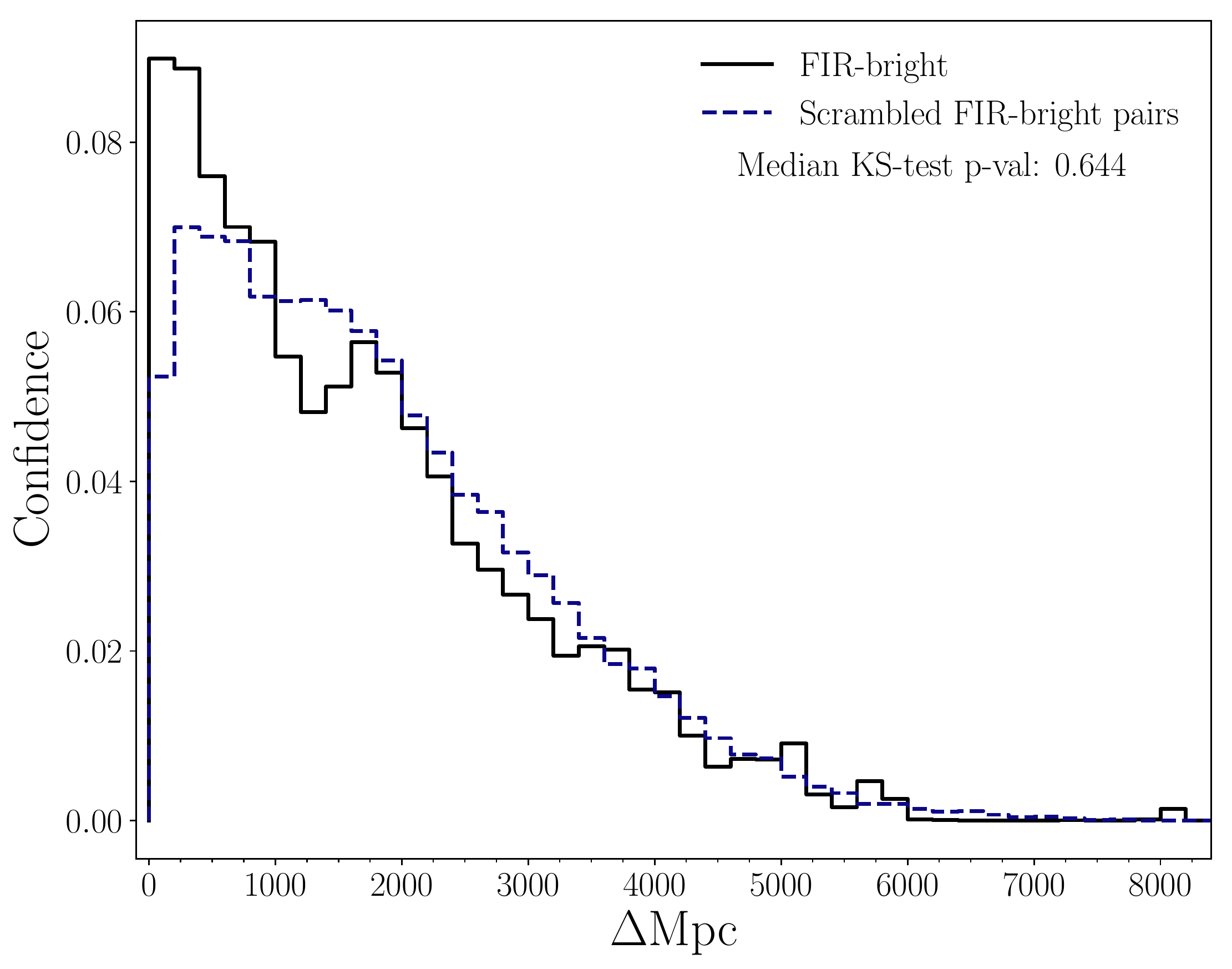}
\caption{The probability of any two pairs being separated by a given \dmpc. In black, we show the distribution of \dmpc~across all 281 FIR sources with more than one bright counterpart (FIR-bright). In dashed purple we show the results of a random resampling of the FIR-bright sample. From the FIR-bright sample, we randomly select 280 sets of 3 counterparts, which will not be physically associated. We calculate their \dmpc, and perform a KS-test on the resampled and the FIR-bright sample to determine if the null hypothesis, that the two samples are drawn from the same parent distribution, can be excluded. This process is repeated 1000 times. We show the median KS-test p-value in the top corner: 0.644. The purple dashed curve shows the normalized results of all 1000 resamplings. These curves have been normalized so that the cumulative probability sums to 1.0.}
\label{fig:dzpdf_all}
\end{center}
\end{figure}

As with Figure \ref{fig:zpdfall}, in Figure \ref{fig:dzpdf_all} we show the cumulative $\Delta$Mpc histogram. This histogram illustrates the difference, in Mpc, along the line of sight for our FIR-bright sample of counterparts. 
For each $\Delta$Mpc value, we find the sum of the probabilities across all FIR objects that found that $\Delta$Mpc as a solution.
Figure \ref{fig:dzpdf_all} therefore illustrates the probability of any two pairs in our FIR-bright sample to be found at a specific $\Delta$Mpc. 
We aim to determine if there is a preferential clustering among the FIR-bright sample. However, comparing to the COSMOS photometric catalogue will be difficult, as we have shown in Figure \ref{fig:zpdfall} that the redshift distributions are different in the two samples, and we will be more likely to find counterparts at smaller \dmpc~separations simply because the redshift peak for the FIR-bright sample is narrow.

We therefore test for preferential clustering in the FIR-bright sample by randomly resampling from the FIR-bright sample itself. If there is a preference for small \dmpc~solutions in the data, randomly resampling should remove it. We therefore assemble all counterpart PDFs, and randomly sample sets of 3 PDFs. We then calculate the \dmpc~values for these three. This is repeated 280 times to replicate the size of the FIR-bright sample, and a cumulative histogram is generated in the same way as for the FIR-bright sample. We use a Kolmogorov-Smirnov (KS) test on the cumulative histograms to assess if the null hypothesis, that the two samples are drawn from the same parent population, can be excluded. We then repeat this random resampling process 1000 times, and perform the KS test for each resampling. We are unable to exclude the null hypothesis at $>3 \sigma$ in 100 per cent of our tests. The median KS-test p-val is 0.644. We interpret this to mean that we are not seeing a statistically significant preferential clustering signal within the FIR-bright sample, and that our data are consistent with being drawn randomly from the FIR-bright redshift distribution shown in Figure \ref{fig:zpdfall}.

\subsection{Probability of consistent redshift}\label{sec:consistentprob}
Thus far, we have focused on the distinctness of the FIR-bright population as a whole, and found that averaged over all of our FIR sources, the contributing counterparts are found to be part of a redshift distribution which is strongly peaked at a redshift of $\sim 1$. However, we have not yet tackled the key question of whether the counterparts which contribute to a specific FIR detection are likely to be physically associated, i.e., if they are at the \textit{same} redshift.  

We define `consistent redshift' as existing within the same redshift bin, which, due to the SED fitting procedure, has a minimum redshift gridding of \dz=0.01. We use the full probability density function to construct a joint probability distribution between all combinations of counterparts associated with a given FIR object, using a methodology described by Equation \ref{eq:jointprob} below.

\begin{equation}
P\left(z_a\text{=}z_b\right) \text{=} \sum_z{ P_a\left(z\right) \times P_b\left(z\right)}
\label{eq:jointprob}
\end{equation}

For each pair of FIR-bright components ($a$ \& $b$), we multiply their probabilities ($P_a$($z$) \& $P_b$($z$)) of existing at a given redshift ($z$), and sum across all redshift solutions. This produces the overall joint probability distribution $P\left(z_a\text{=}z_b\right)$, which is the likelihood that the redshift solutions for components $a$ \& $b$ are within the same redshift bin.
With a larger overlap in the PDFs, or a higher probability of a given redshift solution in the overlapping region, the likelihood that components $a$ \& $b$ are found at consistent redshifts increases.

This calculation is used to assess the plausibility of physical association for any two component pairs blended beneath a given FIR object. If the joint probability is very low (or zero), then there is very little (or no) chance that the two components are physically associated, but rather are line-of-sight projections near to each other on the sky\footnote{We have tested an alternative method of convolving the two redshift probability density functions using test Gaussian distributions. We find that integrating the probability distribution of $z_{a}-z_{b}$, created by convolving the two test distributions, between $\pm\Delta{z}$ is equivalent to Equation 3 so long as the sampling and normalization is done consistently. As Equation 3 is a simpler calculation we proceed with the methodology described above.}.  

If there is no overlap in the PDF solutions at any redshift, the joint probability is 0. This is the case for the PDFs presented in the top panel of Figure \ref{fig:zpdf}; all pairs of components have 0 probability of being found within \dz=0.01.
At the other extreme, if two spectroscopic redshifts are present within the same \dz=$0.01$ bin, the joint probability distribution would indicate that pair has a very high (100 per cent) probability of being found at a consistent redshift. These sources are \textit{potentially} physically associated, and could be in the early stages of a merger. However, we note that at a redshift of 1.0, a \dz~of 0.01 is still probing $\sim$80 Mpc, and so even sources that are found with extremely high probability of being at a consistent redshift are not guaranteed to be physically associated.

\begin{figure*}
\begin{center}
\includegraphics[width=450px]{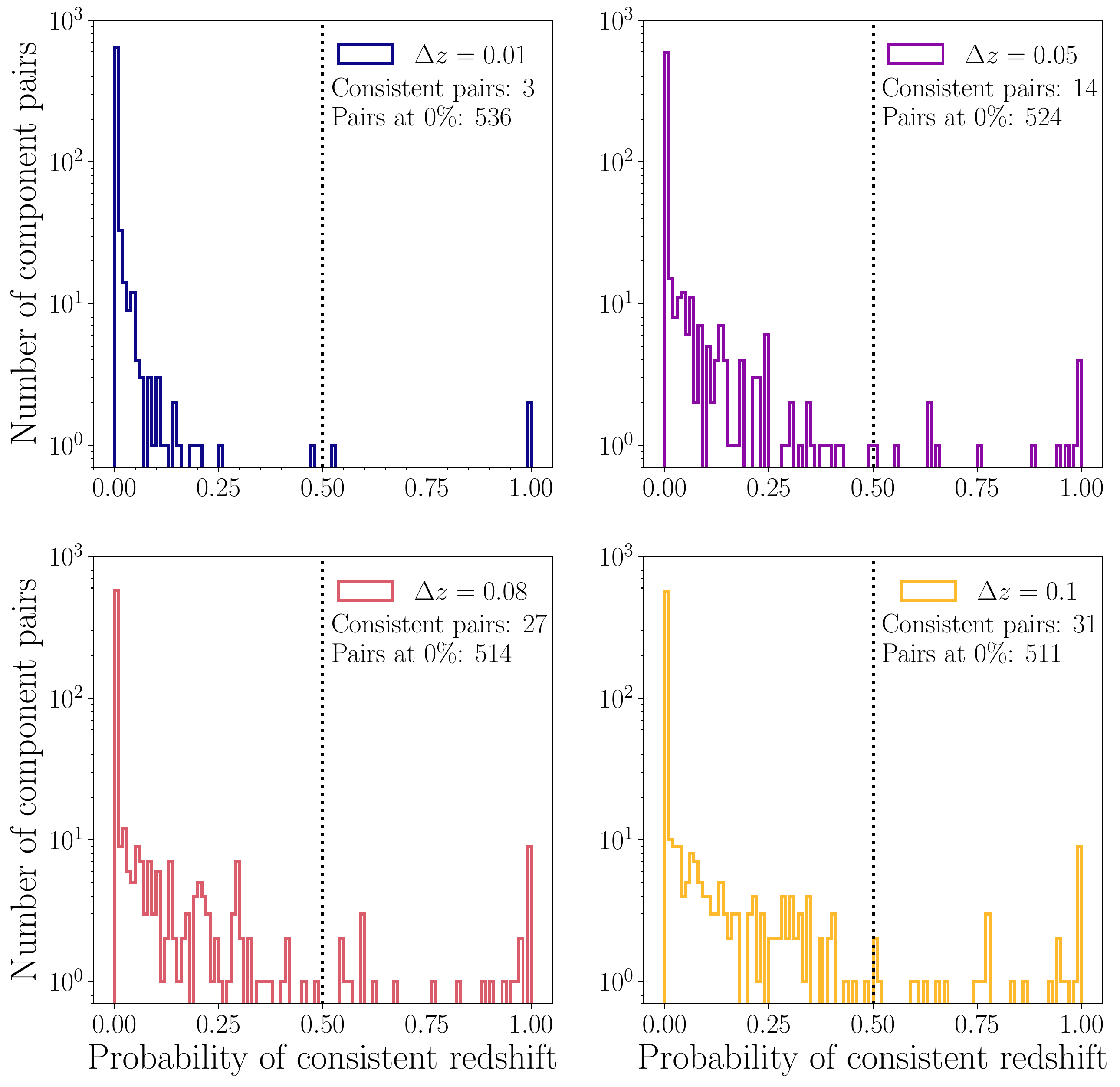}
\caption{The probability of any two component pairs being found within our tested redshift tolerances, \dz $<0.01$, 0.05, 0.08, and 0.1. This figure reflects the distribution of \dz~across all 280 FIR sources with more than one bright, redshift-matched, counterpart (736 counterpart pair combinations). The black dotted vertical line indicates 50 per cent probability. The top left panel shows our fiducial value, of \dz =0.01. The vast majority (99.6 per cent in this case) of FIR sources do not have multiple contributing sources at a consistent redshift. 3 pairs are found with > 50 per cent probability of being found at the same redshift, while 536 pairs have 0 per cent probability of being found at the same redshift. In the top right panel, we expand our redshift tolerance to \dz=0.05, and find 14 (1.9 per cent) plausibly consistent pairs, and 524 (71.2 per cent) counterparts which remain at zero probability of being at a consistent redshift. In the lower left and right panels, we show \dz = 0.08 and \dz =0.1 respectively. Even at these much larger redshift tolerances, the consistent pair fraction remains very low, at 3.7 per cent and 4.2 per cent respectively.}
\label{fig:consistent_all}
\end{center}
\end{figure*}

We compute the joint probability of all 736 counterpart pair permutations, for all fields in which there are $\geq$2 bright components with redshifts. We plot a histogram of the resultant probabilities in the top left panel of Figure \ref{fig:consistent_all}. Three pair combinations (0.4 per cent) are found to have more than 50 per cent probability of being at a consistent redshift. 
The remainder of the sources all have lower probabilities, with the vast majority of the sample found to have no overlap in the PDFs whatsoever. The median value for this histogram is 0, and 536 pair combinations (72.8 per cent) have this value. This distribution indicates that the vast majority of our sample is not comprised of physical pairs.

Because of the discrete nature of the PDFs, it is possible that two counterparts lie within \dz=0.01, but are classed in adjacent bins. This would artificially lower their joint probability. To account for this potential problem, we
 have repeated this analysis with broader redshift tolerances (\dz=0.05, \dz=0.08 \& \dz =0.1). By increasing the redshift bin size, we should capture those counterparts which are close in redshift but not placed in the same bin. We  find that this does not significantly alter our results. The results for these additional three tolerance thresholds are plotted in Figure \ref{fig:consistent_all}. The \dz=0.05 bin width (top right panel) results in 14 consistent pairings, or 1.90 per cent. A \dz =0.08 bin width (bottom left panel) results in a consistency fraction of 3.67 per cent, and \dz=0.1 (bottom right panel) results in a consistent redshift fraction of 4.21 per cent of the sample. The fraction of the sample at a joint probability value of zero remains very high as the redshift bin width increases, which indicates that our data analysis is not substantially affected by well-constrained redshift PDFs in adjacent redshift bins.
 
 We note that we do not expect redshift uncertainties to dramatically alter these results, as the width of the photometric redshift PDFs are taken into account by Equation \ref{eq:jointprob}. If the width of the photometric redshift PDFs are underestimated, then these estimates could also be underestimated, however, by artificially increasing the bin size, we have accounted for galaxies which could be within a \dz~range of up to 0.1 (many hundreds of Mpc). Increasing the bin size will only affect those galaxies with slightly overlapping PDFs or well-constrained PDFs that are very close in redshift; for those galaxies with PDFs which have very different redshift solutions (i.e., those galaxies with 0 probability density in common), increasing coarseness of the PDF by a factor of 10 will not impact the results presented here.

\section{Discussion}	
\label{sec:discussion}

Our results show that the vast majority of the blended FIR-bright counterparts to 24- and 3.6-\mm~selected FIR objects are physically unrelated. A significant portion of our FIR-bright counterparts have no overlap in their redshift PDFs whatsoever, and appear to be consistent with being drawn randomly from the redshift distribution of our FIR-bright sample. Given the longstanding debate on the merger origin of these FIR objects, these results may seem surprising. While the results presented here do not rule out the existence of FIR-bright pairs or mergers in some fraction of the sub-mm population, we suggest that those pairs are not a significant fraction of the 24-\mm-selected sub-mm population\footnote{It is unclear if the population traced by 24 \mm~flux is an unusual subset of FIR detections. We defer a comparison of the 3.6 \mm~FIR-bright counterparts and the 24-\mm~FIR-bright counterparts to a future work.}. We note that this study is fundamentally limited by the resolution of the IRAC 3.6 \mm~FWHM (2 arcsec), and so any interactions at the final stages of a merger will be unresolved. We are furthermore insensitive to any interactions where the companion is FIR-faint or very low mass. However, in low-redshift studies, equal mass encounters are responsible for the strongest starbursts \citep[e.g.,][]{Woods2007, Scudder2012b} and these encounters typically result in roughly symmetric responses in both galaxies \citep[e.g.,][]{Torrey2012, Scudder2012b}. We discuss the resolution limitation further below.  

It is possible that during our redshift identification phase, due to the incompleteness of the redshift matching, we have excluded physical pairs from our analysis. To quantify the extent to which this could affect our results, we estimate how strongly this exclusion could bias the numbers we present here. Our sample has excluded 129 counterparts which did not have available redshifts (see Table \ref{tab:missingz}). We subdivide the sample which is missing redshifts by the number of remaining counterparts which did have redshifts, for that same FIR object. The 37 counterparts which had no redshifts remaining triggered the exclusion of 16 of FIR objects from our analysis. It is unlikely that any bias comes from these missing FIR objects, as no redshifts were preferentially excluded.  However, there was a population of 92 counterparts which were unmatched, for FIR objects with some identified redshifts.

These 92 counterparts represent 11 per cent of the FIR-bright sample. For the 49 counterparts that left only one redshift identification associated with a given FIR object, the failure to match reduced the FIR object count by 38. If we assume that all 38 of the missing FIR objects contained a consistent redshift, and that all 43 of the missing counterparts which did not remove the field from the sample would have given a consistent redshift to one of the remaining counterparts, we can estimate how much our redshift incompleteness could change our results. If we increase the total number of consistent pairs (at $>50$ per cent) from 3 to 84, and the number of total counterpart combinations from 736 to 817, we find that our consistency fraction only increases to 10.2 per cent from our fiducial value of 0.4 per cent. We note that assuming that all of the missing redshifts would have produced a consistent redshift is extremely unlikely, given the distribution shown in Figure \ref{fig:consistent_all}, and we expect the true value to be lower. Even with these conservative adjustments, it is safe to say that the overwhelming majority of the FIR-bright objects are not physically associated.

We also check for AGN contamination in the sample, as \citet{Marchesi2016} showed that the photometric redshifts can be significantly shifted if a strong AGN is present and not accounted for. We cross match with the \citet{Marchesi2016} sample and find that only 18 of our 768 FIR-bright counterparts are within their sample (2.3 per cent).Replacing the PDFs presented here with those calculated with an AGN template does not change the fraction of pairs with $>50$ per cent likelihood of existing at the redshift. At least in this sample, accounting for AGN does not influence these results.

These results mesh extremely well with the newest generation of simulations, which have been treating observational biases more robustly.
As an increasing number of interferometric results have begun to challenge the assumption that the single-dish data is flux coming from a single galaxy \citep[e.g.,][]{Karim2013, Hodge2013, Simpson2015}, simulations are well placed to determine if multiple component systems are likely to be physically interacting or rather are a consequence of projection effects. The current theoretical consensus seems to be that the vast majority of the sub-mm bright sources that are comprised of multiple components should be made up of unrelated components \citep{Hayward2013, Cowley2015, Munoz2015, Cowley2016, Bethermin2017}, with the majority of them suggesting that $\sim70$ per cent of the FIR-bright components should be physically unassociated. While some of these works find a higher median redshift in their simulations \citep{Cowley2015, Munoz2015,Cowley2016} than our data, these simulations were focused on an 850 \mm~selection, which should select higher redshifts, as seen in the observations \citep[e.g.,][]{Casey2012}.  Both \citet{Hayward2013} and \citet{Bethermin2017} find similar typical redshifts. \citet{Cowley2015} suggests that the median difference in redshift between bright components is approximately $\Delta z$ = $1.0$. In particular, \citet{Bethermin2017} finds that only 5 per cent of the blended FIR flux should come from additional components at consistent redshifts, similar to the results presented here.

Both the results from the theoretical works and the results presented here will influence other metrics used to understand the sub-mm/FIR population; e.g., the number counts\footnote{A deeper analysis of the influence of how these results align with the extensive literature on number count predictions and observations is beyond the scope of this work.} and luminosity functions determined by observational works. Furthermore, \citet{Cowley2017} suggests that such blending of unrelated sources will inflate the clustering measurements by a factor of $\sim 4$ for a $\sim 15$ arcsec beam, or more if the beam size is larger. Such clustering measurements (i.e., assessing the overabundance of sources along a line of sight where a FIR source is present), have previously been used to argue in favour of the merger origin hypothesis \citep{Ivison2000, Ivison2007}. We note that in addition to the overestimation of the clustering predicted by \citet{Cowley2017}, very large scale clustering could contribute to the signal seen, while still being at too large a scale to imply gravitational interactions (e.g., Figure \ref{fig:dzpdf_all}).

The vast majority of existing observational works examining the multi-component nature of bright FIR/sub-mm sources do not have redshift measurements \citep[e.g.,][]{Karim2013, Hodge2013, Koprowski2014, Simpson2015}. These studies have instead used statistical arguments based on (e.g.,) an excess of bright counterparts along the line of sight relative to what would be expected from a blank field \citep[see also the Appendix]{Simpson2015}. This lack of redshift information means that there are only a few studies to which we can directly compare our results. 
The most similar work to that presented here is that of \citet{Simpson2016}, who resolve 52 sources contributing to 30 bright SMGs, also selected to have 24-\mm~detections. 
In 11/30 sources, there are photometric redshifts presented for more than one component. Of those, 6 have contributing counterparts which are at inconsistent redshifts (considering the errors on each measurement). 
Another 4 have plausibly overlapping photo-z measurements (within errors), and 1 pair of components has identical measured redshifts. Without the full PDF, it is impossible to say how probable a consistent redshift measurement is for the redshifts which have overlapping error bars, as was done in this work. If we assume conservatively that all of them have >50 per cent likelihood of being found at a consistent redshift, we estimate that over half (54.5 per cent) of the \citet{Simpson2016} photo-z matched sample has incompatible redshift estimators. While this 50 per cent estimate is significantly lower than our current estimate, the selection criteria are very different in our work to the \citet{Simpson2016} work, and 54.5 per cent is a lower, conservative estimate to the incompatible redshift fraction. In both our work and in \citet{Simpson2016}, it is fair to say that a majority of SMGs that are found to have multiple components contributing are comprised of physically unrelated galaxies.

Another recent work, \citet{Hayward2018} follows up interferometric ALMA observations of a small sample of 850 \mm-selected sources which were found to have multiple FIR-bright components, finding redshifts for 9 of those 11. 6/9 sources (67 per cent) are determined to be at inconsistent redshifts, with large error on the percentages due to the small number statistics involved. The conclusions drawn here are entirely consistent with that of the \citet{Hayward2018} work, as they conclude that line-of-sight projections must be common in the data.

Similarly, \citet{Stach2018} uses a combination of photometric redshifts with high resolution ALMA imaging, very similar to our methodolgy. 44 per cent of their sample is found to have multiple bright components, and of those, \citet{Stach2018} reports an excess of \dz<0.25 counterparts, with 24/46 components found at this redshift, concluding that 30 per cent of the sample should be comprised of physical associations. While this is considerably higher than the fraction reported here, \citet{Stach2018} uses much larger redshift tolerance in considering multiple counterparts to be at the same redshift, at \dz=0.25 vs our widest bin of \dz =0.1, which would naturally lead to a higher consistency fraction.

\citet{Wardlow2018} presents the results of a CO(3--2) line search with ALMA for 6 single-dish detections, resolved into 14 bright counterparts. Consistent with what we present here, they find that 83 per cent of their FIR bright sample are found to be at inconsistent redshifts with other FIR-bright counterparts. They conclude, as we do, that the FIR luminous objects are unlikely to be physically and gravitationally bound systems.

The other statistical work involving redshifts for a sub-mm sample is that of \citet{Chapman2005}, which obtained 76 redshifts for a sample of sub-mm galaxies. However, these sources were selected from single-dish observations, so very few of them were already resolved into multiple flux-emitting components. \citet{Barger2012} has only 3 sources which divide into multiple components; one of them has spectra for more than one of those components \citep{Chapman2005, Barger2008}, and they are found at inconsistent redshifts (including error bar uncertainties). Other works have similarly struggled with number statistics for their multi-component systems; both \citet{Smolcic2012} and \citet{Miettinen2015a} have incompatible redshift fractions of 40 and 50 per cent, with samples of 5 and 6 respectively. Similarly, \citet{Danielson2017} report redshifts for a set of 52 SMGs taken from the ALESS sample, but of those, only three are found to have reliable redshifts for multiple components. Of those, they report no evidence for clustering along the line of sight, with redshift differences ranging from \dz =0.06--1.25. These results are entirely consistent with the picture that we present here, of strong contamination along the line of sight when multi-component systems are present.

We note that the chain of thought which led to the broad assumption that many FIR/sub-mm sources are of a merger origin is an interesting one; we refer the interested reader to the Appendix, where we have undertaken a more detailed discussion of this point.

As mentioned earlier in this section, the current study is limited in scope to only sources which are spatially resolved in the 3.6-\mm~data, i.e., typically separated by at least 2 arcsec\footnote{At our typical redshift of $z$=1, this is a separation of 16.2 kpc \citep{Wright2006}}. We are not sensitive to any merging populations which are in their final stages, or, as some studies have observed, multi-nucleus systems at sub-arcsecond resolutions \citep[e.g.,][]{Ivison1998, Smail2003, Alaghband2012, Menendez2013}. Indeed, a recent study of a local Ultra-Luminous Infrared Galaxy (ULIRG\footnote{ULIRGs are defined to have IR luminosity L$_{IR}$ > $10^{12}$ L$_{\odot}$}) triple system found that if it were to be found at high redshift, it would likely be misidentified as a clumpy galaxy or a small group \citep{Vaisanen2017}. However, our results hold for any well resolved components at arcsecond scales, such as those found by \citet{Karim2013}, where the typical separation is $\sim6$ arcsec. At our typical redshift, 6 arcsec corresponds to a physical scale of $\sim$ 49 kpc \citep{Wright2006}, a scale at which galaxy--galaxy interactions are seen to impact the star formation rates of galaxies in the local universe \citep{Scudder2012b, Patton2013, Ellison2013}. Furthermore, if one object in a pair is not FIR-bright, it would not be present in our sample, as we only consider sources which are FIR-bright. However, in local mergers, both galaxies in a major (i.e., roughly equal mass) merger are expected to be equally affected by the interaction and both should show significant star formation \citep[e.g.,][]{Torrey2012}.

We therefore propose that the FIR population in the COSMOS field, in particular the subset traced by 24-\mm~flux, is dominated by multiple flux-emitting components which are found at inconsistent redshifts. These galaxies are therefore not physically interacting, but are line-of-sight projections. Such projections are a well understood source of contamination in low-redshift galaxy pairs samples \citep[e.g.,][]{Patton2008} and so it is perhaps unsurprising that given the large range of redshift visibility that the sub-mm bright population enjoys, that there is strong contamination along the line of sight.
There remains a small fraction of these systems that are found at consistent redshifts both in the current work and in the literature \citep[e.g.,][]{Tacconi2006}. In this work we do not intend to entirely rule out a merger origin for some fraction of the FIR-bright sources. We also recognize that some sources have been found to lie in protocluster environments \citep{Hodge2013b, Ivison2013}. However, considering that the vast majority of our \herschel~sources have spectro/photometric redshift PDFs that do not overlap at all, we must conclude that the majority of these galaxies are entirely gravitationally unrelated. 

We therefore suggest that while the FIR population is more tightly clustered in redshift space than the overall distribution of photometric galaxies in the COSMOS field, this population is constructed of galaxies that are not physically associated, and the blending of their FIR light within the beam of \herschel~has resulted in a boost to the detected FIR luminosity. We suggest that any merging population is either present at smaller spatial scales than we are sensitive to with 3.6-\mm~resolution, that SF has been triggered in these galaxies in a very asymmetric fashion, or that the SF in these galaxies is driven by non-merging processes. These processes would not need to produce star formation rates as extreme as originally proposed, when all FIR flux had been assigned to a single physical system. Dividing the flux among 2--3 unrelated line-of-sight systems means that the SFR along the line of sight, while significant, is a considerable overestimation of the SFR in a physically bound system. The nature of the individual FIR-bright objects is impossible to constrain using the current data, and is likely to be a mixture of late-stage mergers, secular star formation, and other systems.

\section{Conclusions}
\label{sec:conclusions} 
We briefly summarise the findings of our work here. We have investigated the physical association between a sample of FIR-bright (30 mJy $\lesssim$ FIR flux $\lesssim$ 110 mJy), blended sources in the COSMOS field. Our sample is selected to have both 3.6-\mm~and 24-\mm~counterparts, and in a previous work \citep{Scudder2016} we used the {\sc{xid+}} software to assign best-fitting fluxes to each counterpart through Bayesian inference methods. In this work we use only the sources identified as FIR-bright to investigate how close in redshift these contributing counterparts are.
\begin{itemize}
\item We select only those sources that were identified to be contributing more than 10 per cent of the total FIR source flux, and cross-match them with spectroscopic \citep{Davies2015} and photometric \citep{Laigle2016} redshift catalogues. We have a success rate in our crossmatching of 86 per cent, and the majority of the missing sources are near bright stars, which has contaminated their optical photometry.
\item We extract those FIR objects that have more than one bright component, and where more than one counterpart redshift has been identified. This results in a sample of 280 FIR objects, divided into 768 bright counterparts, with 736 pair permutations.
\item We find that the FIR-bright subsample is more densely clustered between 0.5 $\leq z \leq$ 1.5 than the photometric catalogue of \citet{Laigle2016}. However, the FIR-bright sample appears to be consistent with a random sampling of this narrower redshift distribution, as KS-tests on a scrambled version of the FIR-bright sample is unable to reject the null hypothesis of being drawn from the same parent population at $>3\sigma$.
\item We find that 72 per cent (536 pairs) of the FIR-bright sample has no overlap in their redshift distributions whatsoever, indicating 0 probability that they are found at a consistent (\dz =0.01) redshift. Only $0.4$ per cent of the sample (3 individual pairs) is found to have more than 50 per cent likelihood of existing at consistent redshifts (where consistent is \dz $<0.01$). Increasing the redshift tolerance does not substantially change these results.
\end{itemize}

In our sample, potentially interacting FIR counterparts comprise a minority of the overall population. These results caution that future studies of the sub-mm galaxy population require redshift estimations to be made for all counterparts before any assumptions on physical associations can be made.

\section*{Acknowledgments}
We thank the anonymous referee for a careful and thoughtful reading of this manuscript, and for their feedback which helped improve the presentation of this work.

JMS thanks Michael Zemcov, Andreas Papadopoulos, Steven Duivenvoorden, Gianfranco De Zotti, Mattia Vaccari, and Mathieu B\'ethermin for useful comments on a draft of this manuscript, which improved the clarity of this work.

JMS and SJO acknowledge support from the Science and Technology Facilities Council (grant numbers ST/L000652/1).

JLW acknowledges the support of a STFC Ernest Rutherford Fellowship and has received additional support from a European Union COFUND/Durham Junior Research Fellowship under EU grant agreement number 609412, and STFC (ST/P004784/1)

This research made use of Astropy (\href{http://www.astropy.org}{http://www.astropy.org}), a community-developed core Python package for Astronomy \citep{astropy2013}, the matplotlib python plotting package \citep{matplotlib}, and numpy for array-like data manipulation \citep{numpy}.

The research leading to these results has received funding from the Cooperation Programme (Space) of the European Union's Seventh Framework Programme FP7/2007-2013/ under REA grant agreement number [607254].

The {\it Herschel} spacecraft was designed, built, tested, and launched under a contract to ESA managed by the Herschel/Planck Project team by an industrial consortium under the overall responsibility of the prime contractor Thales Alenia Space (Cannes), and including Astrium (Friedrichshafen) responsible for the payload module and for system testing at spacecraft level, Thales Alenia Space (Turin) responsible for the service module, and Astrium (Toulouse) responsible for the telescope, with in excess of a hundred subcontractors.

GAMA is a joint European-Australasian project based around a spectroscopic campaign using the Anglo-Australian Telescope. The GAMA input catalogue is based on data taken from the Sloan Digital Sky Survey and the UKIRT Infrared Deep Sky Survey. Complementary imaging of the GAMA regions is being obtained by a number of independent survey programmes including GALEX MIS, VST KiDS, VISTA VIKING, WISE, Herschel-ATLAS, GMRT and ASKAP providing UV to radio coverage. GAMA is funded by the STFC (UK), the ARC (Australia), the AAO, and the participating institutions. The GAMA website is http://www.gama-survey.org/ . 

This work is based on data products from observations made with ESO Telescopes at the La
Silla Paranal Observatory under ESO programme ID 179.A-2005 and on
data products produced by TERAPIX and the Cambridge Astronomy Survey
Unit on behalf of the UltraVISTA consortium.

\bibliographystyle{mnras}
\bibliography{test_imac.bib} 

\appendix
\label{section:appendix}
\subsection{Previous observational results}

The historical assumption that many sub-mm galaxies are interactions grew from a series of natural assumptions; firstly, that they drew obvious parallels with the ULIRG population at low redshift, in that they were similarly rare in their volumes, and both were very IR luminous. ULIRGs are almost entirely merging systems \citep{Sanders1996}, so the interpretation of high-z IR bright systems as the high-z extension of the ULIRG population was a straightforward one.
However, testing this merger-induced star formation hypothesis observationally has proven to be a particular challenge, as it requires high resolution data to both obtain accurate redshift estimations and accurately determine counterparts. 

Obtaining any redshifts in the first place proved to be a considerable challenge, as sub-mm sources either had no multi-wavelength counterpart \citep{Frayer2000, Dannerbauer2008} or had a number of potentially contributing counterparts \citep{Hughes1998, Hatsukade2010}, and in either case, photometric redshifts were particularly badly constrained. Of the 5 SMG sources, \citet{Hughes1998} found somewhere between 2 and 7 feasible counterparts, each with their own, generally poorly constrained, redshift estimates. Other work from this era struggled with the same problems of small number statistics and an overabundance of possible counterparts \citep[e.g.,][]{Downes1999}. 

Where redshifts could be obtained, a small number of these galaxies were observed in high resolution with  (e.g.,) the Plateau de Bure interferometer or the Very Large Array. These samples typically presented new observations of only 1 -- 3 systems \citep{Ledlow2002, Neri2003, Greve2005, Hainline2006, Tacconi2006, Tacconi2008, Engel2010, Ivison2010, Bothwell2010, Aravena2010} and found that these systems were very irregular in their morphologies, showing either clumpy components, irregular line profiles, or extremely high gas densities. \citet{Greve2005} notes that with the sorts of gas densities being observed in these systems, the galaxies should fail the Toomre gas stability Q parameter \citep{Toomre1964}, and be extremely prone to collapse, either through disk fragmentation, or through merger-induced instabilities. Many of these works interpret the disorder in the line profiles or the gas density measurements as the signatures of merging events \citet{Tacconi2008, Tacconi2006, Engel2010}.
It is not particularly clear whether the high gas surface densities should be due to interaction-driven tidal torquing, as this torque has been shown to be relatively ineffective in very gas rich systems \citep{Hopkins2009}. The torquing mechanism requires the presence of a considerable stellar bar, or the gas inflow to the center of the system becomes extremely inefficient.

Individual clumps have also been taken as signs of interactions, or pre-coalescence mergers \citep{Smail2003, Nesvadba2007}.  Generally, asymmetric or ``messy'' morphologies are attributed to the influence of interactions between galaxies \citep{Younger2010, Engel2010, Menendez2013, Chen2015}. At low redshift, such asymmetries are good tracers of gravitational interactions between galaxies.
The clumpy nature of high redshift galaxies on small scales has also become considerably more difficult to interpret; studies like that of \citet{Law2007} and \citet{Cibinel2015} show that morphological irregularities do not seem to correlate with any other properties of the galaxy, and so these asymmetric morphologies or multiple components in high redshift systems are either not good proxies of an interacting system, or the interaction is not changing the observational properties of the galaxy.  \citet{Law2007} specifically find that the sub-mm population of \citet{Chapman2005} is no more likely than an isolated sample to show multiple strong nuclei. Similar results were found in \citet{Swinbank2010}, with no excess of asymmetry found in the SMG sample. Some works are beginning to conclude that some intermediate redshift galaxies are very messy disks fuelled by pristine gas infall \citep[e.g.,][]{Carilli2010}, which explains the ordered nature of their rotation. \citet{Nayyeri2017} finds that for a single lensed galaxy, the SFR is consistent with the typical SFR for galaxies of that stellar mass at a redshift of 2.6.

Studies that were more statistical in nature found that the overabundance of possible counterparts persisted beyond the brightest sources. Such studies either relied on radio source counterparts to identify a \textit{single} "true" counterpart IDs \citep[e.g.,][]{Ivison2007}, or found the optical counterpart that aligned most closely with the centroid of the sub-mm detection. (The exception to this was if the closest galaxy appeared to be a giant elliptical without much star formation, whereupon gravitational lensing would be suspected of a fainter, bluer object.) These studies still often found multiple possible counterparts to the FIR flux, and the interpretation of this statistical overabundance was explained using a probabilistic argument. These sub-mm counterparts had more possible counterparts along the line of sight than could be expected by randomly sampling the field; this led many works to the conclusion that these counterparts must be physically associated \citep{Ivison2000, Ivison2007}. \citet{Hodge2013} further noted that while they find an excess of sources in the majority of these fields with ALMA observations, there is no excess of sources at the smallest separations, which is suggested to be due to a merger origin.

However, the true counterparts must always be identified in the sub-mm in order to be conclusive, and prior to ALMA, this meant that many works were conducted with the SMA. However, due to the longer observing times required, this usually resulted in samples that were rather small \citep{Iono2006, Hatsukade2010, Younger2007, Younger2008, Younger2009, Younger2010} sensitive to the brightest counterparts, and primarily used for the testing of the reliability of multi-wavelength counterpart identification. 

Our understanding of multiplicity took another turn with the advent of high resolution ALMA follow-up of single-dish data, where it was discovered that a considerable fraction of the sub-mm detected objects were resolved into multiple flux-emitting components \citep[e.g.,][though cf \citealt{Koprowski2014}]{Karim2013, Hodge2013, Simpson2015}. These surveys, however, had no intrinsic redshift measurements associated with any of the individual flux-emitting components, as they are continuum studies, and the sub-mm population is well known to be luminous over a wide range of redshifts. However, the statistical argument of the overabundance of components along the line of sight to a sub-mm source implying a physical association (i.e., interactions) persisted. \citep{Simpson2015}.

Without measurements of the spectroscopic or photometric redshifts, this assessment of the overabundance of flux-emitting components as physically associated is impossible to ascertain directly. Existing studies of the redshift distributions of the resolved sub-mm galaxy population are still relatively limited in statistical power \citep[e.g.,][]{Hatsukade2010, Barger2012, Smolcic2012a}, or contain a number of poorly constrained photometric redshifts, where the range in possible redshifts is sufficiently broad that it is impossible to rule out a consistent solution \citep[e.g.,][]{Younger2009}, though we note that the best fit solutions for the galaxies in \citet{Younger2009} are widely separated in redshift space. \citet{Smolcic2012} presented interferometric imaging of a sample of 19 LABOCA-selected galaxies, where several of the galaxies had photometric redshift information. Of the 5 with multiple components and multiple redshifts, 2 must be at inconsistent redshifts; the remaining 3 are potentially consistent within the very wide error bars presented.  \citet{Miettinen2015a} also presented redshift information for a subset of the multiple component systems; of the 6 systems with more than one redshift associated, 3 of them have inconsistent redshifts, and the other 3 are presented only as lower limits to the redshift estimate, so it is impossible to determine whether the remainder are also inconsistent.

\subsection{Theoretical results}

Simulations of the high redshift galaxy population have historically struggled to reproduce the extreme infrared luminosities required to explain the sub-mm galaxy population, and have invoked a number of possible explanations in order to explain the nature of these sub-mm luminous sources, which have largely been interpreted as single galaxies.

With the observational base seemingly convinced that the sub-mm galaxy population was largely constructed of interactions, it should come as no surprise that many of the simulations set about testing this assessment. The most luminous galaxies have often been ascribed to merger induced starbursts \citep{Genel2008, Dekel2009, Narayanan2010, Hayward2011} either through an abundance matching method \citep[e.g.,][]{Genel2008,Dekel2009}, by directly modeling the sub-mm flux through radiative transfer codes such as {\sc{sunrise}} \citep{Jonsson2006,Jonsson2010, Narayanan2010, Hayward2011}; with codes like {\sc{grasil}} \citep{Silva1998, Swinbank2008} the geometry of the system could also be taken into account.
However, even amongst these works, the slightly less FIR-luminous (though still certainly detectable) populations rapidly became heterogeneous, with a number of works suggesting that the less luminous population was a mixture of interacting and non-interacting systems \citep{Narayanan2010, Dave2010, Hayward2011}. Others suggested that these less luminous systems were being fueled by pristine gas infalling all the way to the center of the galaxy's halo \citep{Granato2004, Genel2008, Dekel2009, Narayanan2015}

Other simulations, however, suggested that mergers should not play a major role in triggering the star formation within these luminous sub-mm galaxies. Instead, these gas rich systems are hypothesized to be simply too unstable to survive without large-scale disk fragmentation \citep{Immeli2004, Bournaud2009, Lacey2016}, which would then transform a gas rich disk galaxy into something highly morphologically disturbed, without requiring any kind of external perturbation, like an interaction.  These simulations expect to observations to reveal significant clumps of star formation within the disk, and predict that these clumps are the results of vigorous star formation in systems that are gas-dominated \citep{Bournaud2009} \& without a strong bulge component \citep{Immeli2004}.

The last suggestion used to explain the extreme star formation rates in sub-mm sources has been an alteration of the initial mass function (IMF). A variable IMF is invoked as a way to inflate the IR luminosity without needing to dramatically increase the star formation rate in high redshift systems. If there were a mode of star formation (perhaps triggered by minor interactions) which preferentially formed high mass stars, then the UV luminosity would be increased on average. These high mass stars are also producers of significant volumes of dust, which then provides the mechanism to boost the IR luminosity observed. The dust produced by these stars will be heated by the UV radiation of other massive stars, creating an excess of IR light. The most extrteme of these variations to the IMF was proposed by \citet{Baugh2005} in an early version of the {\sc{galform}} model \citep{Cole2000}. Some element of this IMF variability has persisted in subsequent versions of {\sc{galform}} \citep{Lacey2016}, though the change in the IMF between normal and starburst modes is not quite as extreme.

It has only been recently that the simulations have begun to tackle observational biases in the FIR samples more thoroughly, and it has been since these studies of the biases of single-dish observations that a consensus has emerged that FIR-bright sources in single-dish data are likely to be strongly blended, and that those blends are unlikely to be physical associations of galaxies \citep{Hayward2013, Cowley2015, Munoz2015, Cowley2016, Bethermin2017}.

\bsp	
\label{lastpage}
\end{document}